# A Carbon Dioxide Absorption System Driven by Water Quantity

Xiaoyang Shi,[1] Hang Xiao,[1] Xi Chen,[1*] Klaus. S. Lackner[2*]

[1] Department of Earth and Environmental Engineering, Columbia University, New York, NY 10027, USA

[2] School of Sustainable Engineering & Built Environment, Arizona State University, Tempe, AZ 85287-9309, USA

* Corresponding authors: xichen@columbia.edu (X.C.), Klaus.Lackner@asu.edu

## Abstract

A novel system containing nanoporous materials and carbonate ions is proposed, which is capable to capture $CO_2$ from ambient air simply by controlling the amount of water (humidity) in the system. The system absorbs $CO_2$ from the air when the surrounding is dry, whereas desorbs $CO_2$ when wet. A design of such a $CO_2$ absorption/desorption system is investigated in this paper using molecular dynamics and quantum mechanics simulations, and also verified by experiments. Its working mechanism is revealed as the reduction of free energy of the carbonate ion hydrolysis with the decrease of the number of water molecules in confined nano-pores. The influences of pore size, spacing of cations, surface hydrophobicity and temperature on $CO_2$ capture efficiency are elucidated. The study may help to design an efficient direct air capture system and contribute to the "negative carbon emission".

# 1. Introduction

Ion hydration/dehydrations containing interfaces are ubiquitous and play a key role in a plethora of chemical,[1] physical,[2] biological[3] and environmental systems[4]. Hydration of ions affect not only the structure and dynamics of neighboring water molecules,[5, 6] but also the polarization and charge transfer via the formation of highly structured water molecules.[7, 8] On account of the hydrated water at interface, the rate and extent of various chemical reactions can be greatly improved.[9-12] In essence, ion dehydration gains energy through the release of hydrated water molecules, whose characteristics can be applied to energy storage with anhydrous salts,[13] optimization of protein-drug interaction by releasing localized water,[14] and gas separation via modified water amount of sorbents, among others.[15-18]

Previous experimental observations have gained useful insights into the detailed structural information and dissociation of hydration water of the interface at the molecular level.[19-24] A better understanding of the interaction among ions, ion pairs and solid-liquid interfaces can be revealed by molecular modeling,[25-29] such as evaluating the ion hydration free energy,[30-32] analyzing the water density at solid hydrophobic surfaces,[33, 34] as well as understanding the hydration energy and structural change with reduced water activity,[29, 35, 36] which showed a high degree of positional ordering parallel to the surface. The study on mineral surface indicated that the water adsorption on all calcite surface planes is energetically favorable.[37] An *ab initio* calculation compared the dissociative and associative adsorption water on the calcite surface indicated that water dissociation is disfavored except near a carbonate ion.[38-40] Ion hydration at interface has resulted in a flurry of interest in materials chemistry and physics, and a better understanding of the structure and dynamics of water/solid interface responds to the pressing contemporary issues such as climate change and challenges in water and energy.[41]

Climate change concerns may soon force drastic reductions in $CO_2$ emission.[42] and the urgency of the development of $CO_2$ capture from ambient air was demonstrated elsewhere[43]. Of particular interest in this study is a novel mechanism for direct air capture of $CO_2$ driven by ion hydration/dehydration energy change in a series of nanoporous materials, such as carbon nanotube, activated carbon, graphene aerogel, zeolite, and ion exchange resin (IER)[44-46]. Our experiments have shown that the controllable ion hydrations may drive these nanoporous materials to absorb $CO_2$ when the surrounding atmosphere is dry, and release $CO_2$ when wet. The process of $CO_2$ absorption/desorption can be depicted as a series of reactions of water dissociation, formation of bicarbonate and hydroxide ions, and $CO_2$ combination, as shown Eq. 1-4 and Fig. 1.

$$H_2O \Leftrightarrow H^+ + OH^- \quad (1)$$

$$CO_3^{2-} + H^+ \Leftrightarrow HCO_3^- \quad (2)$$

$$OH^- + CO_2 \Leftrightarrow HCO_3^- \quad (3)$$

$$HCO_3^- + HCO_3^- \Leftrightarrow CO_3^{2-} + CO_2 + H_2O \quad (4)$$

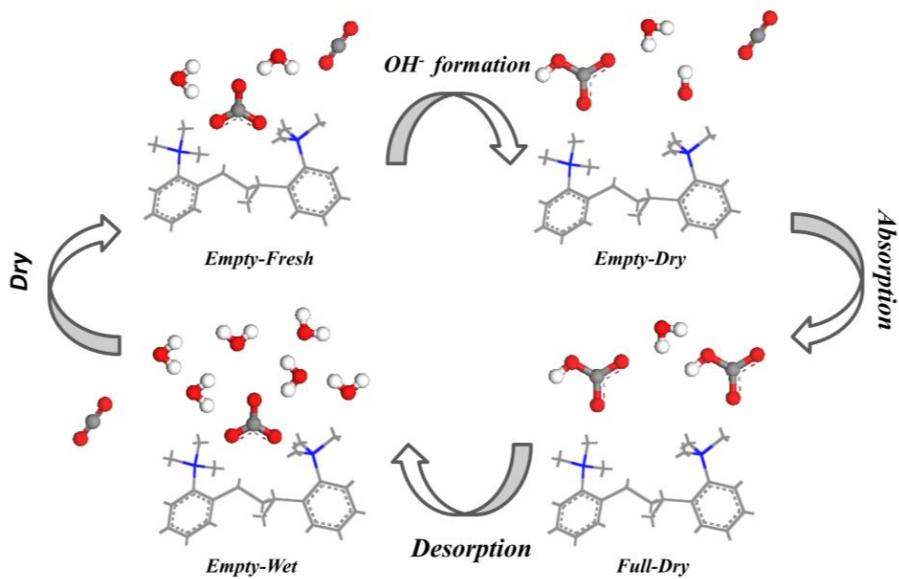

Fig. 1 Reaction pathway of $CO_2$ absorption/desorption on IER. ***Empty-Fresh*** state is the sorbent in dry condition with a few water molecules in the surrounding. ***Empty-Dry*** state is $H_2O$ splits into $H^+$ ion and $OH^-$ ion which is ready to absorb $CO_2$, and $H^+$ ion is combined with $CO_3^{2-}$ forming $HCO_3^-$ ion, Eq. (1) and (2). ***Full-Dry*** state is the full-loaded sorbent in the dry condition, Eq. (3). Eqs. (1)-(3) present the absorption process.

***Empty-Wet*** state is regenerate absorbent releases CO$_2$ in the wet condition (desorption), Eq. (4).

The previous mechanism study of free energy change was based on ions with different number of water molecules in vacuum surroundings[47]. Wide applications of hydration swing to absorb CO$_2$ require a more detailed understanding of the molecular mechanisms of the hydration induced energy change at an ion hydration/solid interface. Using atomistic simulations, the mechanism of CO$_2$ absorption with respect to water quantity may be elucidated via the explorations of the reaction free energy of carbonate ion hydrolysis in a confined nanoenvironment. To optimize the performance of the CO$_2$ capture system, a systematic study needs to be carried out on the efficiency of effective hydration driven CO$_2$ capture effect, with respect materials and system parameters such as different pore sizes, hydrophobic/hydrophilic confined layers, temperatures, and spacing between cations on the absorbent.

The present study employs Molecular Dynamics (MD) and Quantum Mechanics (QM) simulations to reveal the ion hydration energy changes with water numbers under the condition of nano-confined layers, from which the mechanism of hydration driven absorption for CO$_2$ capture from ambient air is explained systematically. Based on the mechanism of humidity-driven CO$_2$ absorption/desorption, the CO$_2$ capture performances are optimized through a parametric study. The study may shed some insights on the future research of low-cost high-efficient CO$_2$ capture system, and contribute to other related areas such as ion hydrations and water/solid chemical reactions.

## 2. Model and Computational Method

### 2.1 Model

A CO$_2$ capture system absorbs more CO$_2$ in a relative dry condition (when the water number per reaction, $n$, is small) with the help of more OH$^-$ ions, while absorbs less CO$_2$ with less OH$^-$ ions in a relative wet condition. According to Figure 1, the reaction pathway of CO$_2$ absorption is:

$$\text{CO}_3^{2-} + n\text{H}_2\text{O} \Leftrightarrow \text{HCO}_3^- + \text{OH}^- + (n-1)\text{H}_2\text{O} \quad (5)$$

This shift in equilibrium appears counter-intuitive going against the mass action law implicit in Eq. 5. The reason is that the driving force to generate OH⁻ ion in equilibrium on the ion hydration/solid surface is the change in the size of ion hydration clouds in the system. This phenomenon cannot be observed in a regular salt solution surrounding, because the ratio of carbonate ion to water molecules, say $CO_3^{2-}$:$H_2O$ is 1:20 in a saturated sodium carbonate solution at 25 °C, is much lower than the one in water vapor/ion surroundings. Therefore, the total equation taking into account hydration water is refined as Eq. 6

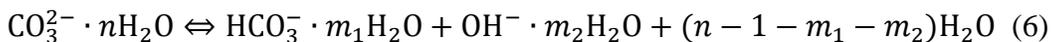
$$CO_3^{2-} \cdot nH_2O \Leftrightarrow HCO_3^- \cdot m_1H_2O + OH^- \cdot m_2H_2O + (n - 1 - m_1 - m_2)H_2O \quad (6)$$

The chemical reaction moves to the right hand side with smaller number of water molecules to produce more OH⁻ ions which is conducive to absorb $CO_2$, and swings to the left hand side with larger number of water molecules. Note that the partial pressure of $CO_2$ over a wet fully-loaded resin, which is at the bicarbonate state, is comparable to the equilibrium partial pressures over sodium bicarbonate brines of approximately one-molar strength.[48] This suggests that the unusual state is not desorption state but the absorption state in $CO_2$ capture system.

Therefore, we postulate that the energetically favorable state of this system can be shifted with the different numbers of hydrated water molecules around ions. A methodology combined with MD and QM is outlined in Fig. 2 to overcome the limitations of MD on simulating bond breaking/forming, whereas full QM or *ab initio* MD would computationally expensive. A sequential molecular process may be established in the corresponding thermodynamic cycle. Let ΔE₁ and ΔE₂ represent the hydration standard-state energy changes of system 1 (S1, a carbonate ion) and system 2 (S2, a hydroxide with a bicarbonate ion), respectively. ΔE₃ represents the standard-state energy change of reaction $CO_3^{2-} + H_2O \Leftrightarrow HCO_3^- + OH^-$ in vacuum at room temperature. ΔE₁ and ΔE₂ can be determined by MD simulations; the state of ΔE₃ can be deduced from QM simulation. The total energy change ΔE=ΔE₁+ΔE₂+ΔE₃ of Eq. 6 can be evaluated as the number of surrounding water molecules (*n*) changes. To balance the anionic charges, freely movable sodium cations (Na⁺) are included in the computational cell. Note that the entropy change is not calculated since its impact on the system is small.

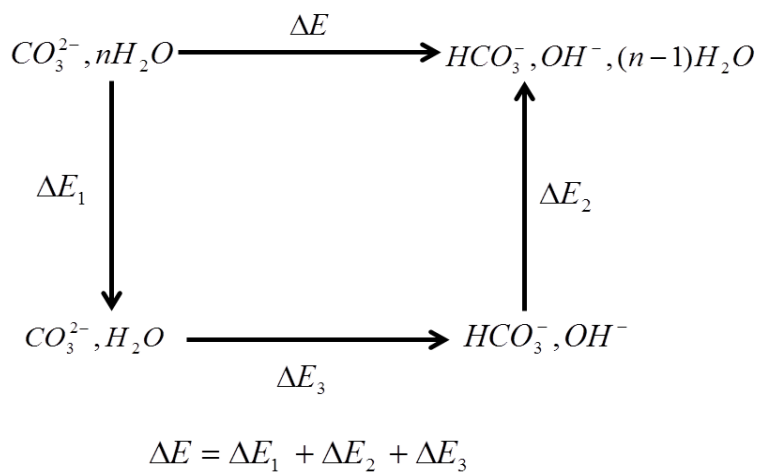

$$\Delta E = \Delta E_1 + \Delta E_2 + \Delta E_3$$

Fig. 2 Thermodynamic cycle of reaction energy change

## 2.2 Computational Cell

The proof-of-concept computational cell consisted of a graphene layer attached 200 sodium cations, and 100 moveable carbonate ions (S1) as reactant, or 100 bicarbonate and 100 hydroxide ions (S2) as product with different number (100 to 1500) of water molecules. A repeated unit configuration is shown in Fig. 3. The graphene was treated as a rigid plate. Periodical boundary conditions (PBCs) were employed in three dimensions.

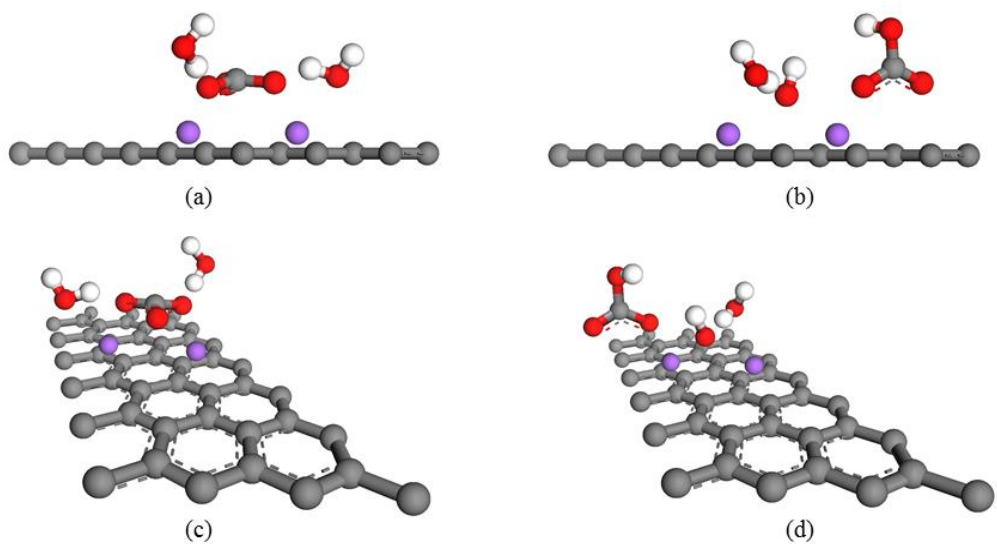

Fig. 3 (a) (b) The computational cell of the model S1 and S2 orthographic lateral view. (c) (d) Model S1 and S2 perspective plane view. Grey skeleton represents graphene, red ball represents oxygen, white ball represents hydrogen, and purple ball represents sodium. Graphene was treated as a rigid plate with fixed sodium cations. The ratio of carbonate ion to water molecules is 1:2 in the figure, which is only one example of various ratios of carbonate ion to water molecules. (1:1, 1:2, 1:3, 1:4, 1:5, 1:6, 1:7, 1:8, 1:9, and 1:15 studied in this paper) The initial distance between sodium cations was 3.5 Angstrom and 14 Angstrom, along x and y direction respectively.

The computational model represents a $CO_2$ capture system driven by humidity. The graphene layer is a representative hydrophobic surfaces. $Na^+$ ions represent a series of cations, (which can be $K^+$ ions and $NH_4^+$ ions or others in practice), attached to the surface of the hydrophobic material, like the ion exchange resin in practice. In this system, the varied environmental factors include water quantities, pore size (space between the surface layers), distance between the attached cations, hydrophobicity of the surface layer, and surrounding temperature, etc., to be studied in the following.

MD simulations were carried out using Materials Studio with a COMPASS forcefield,[49] which is an *ab initio* forcefield optimized for condensed-phase application, assigned to all atoms in graphene, carbonate ion, bicarbonate ion, hydroxide ion, and water. Geometry and partial charges on all atoms of anions in gaseous and aqueous phases were calculated by the $DMol^3$ program.[50] Geometry optimizations and population analysis of the anions were obtained according to GGA HCTH methods and the DNP 3.5. A p-type polarization function was employed for hydrogen bonding. The SPC variable bond water model was used in our model. Minimizations were carried out by Quasi-Newton procedure, where the electrostatic and van der Waals energies were calculated by the Ewald summation method[51] (the Ewald accuracy was 0.001kcal/mol, and the repulsive cutoff for van der Waals interaction was 6 Angstrom). MD simulations for all configurations of systems were performed in a NVT-ensemble (constant-volume/constant-temperature) at 298 K. A time step of 1.0 fs was used in all simulations. In most cases, the equilibrium values of thermodynamic parameters were reached within the first 50 ps for NVT using a Nose/Hoover thermostat. All MD simulations were performed for 200 ps to achieve equilibrium followed by a 300 ps simulation for parameter deduction.

The chemical reaction energy of $Na_2CO_3 + H_2O \Leftrightarrow NaHCO_3 + NaOH$ (S3) in vacuum at ground state was simulated via first principle calculation. $Na_2CO_3$ with $H_2O$, and $NaHCO_3$ with $NaOH$ were treated as reactants and products respectively. The total energy at 0K was obtained via functional GGA HCTH[52] and basis set DNP. Enthalpy correction at finite temperatures was computed according to Hessian evaluation of the translational, rotational and vibrational contributions.

## 3. Results and Discussion

### 3.1 Fundamental mechanism of CO2 capture system driven by water quantity

We first explored the fundamental mechanism of $CO_2$ capture system driven by water quantity. From MD and QM simulations, the energy difference of the system between two states was calculated as the number of water molecules was varied. The energies of system 1 (S1, one layer graphene with two $Na^+$, one $CO_3^{2-}$ and one $H_2O$), and system 2 (S2, one layer graphene with two $Na^+$, one $HCO_3^-$ and with one $OH^-$) were plotted against the variation of the number of water molecules (*n*) in Fig. 4a and Fig. 4b, for energy and enthalpy respectively. In the carbonate ion system (S1), the $CO_3^{2-}$ to water ratio is selected to be 1:1, 1:2, 1:3, 1:4, 1:5, 1:6, 1:7, 1:8, 1:9, and 1:15 respectively, and for the bicarbonate ion system (S2), the $HCO_3^-$ to water ratio is established at 1:0, 1:1, 1:2, 1:3, 1:4, 1:5, 1:6, 1:7, 1:8, and 1:14 respectively, from low to high relative humidity. These cases have one-to-one correspondence because of the reaction between one carbonate ion and one water.

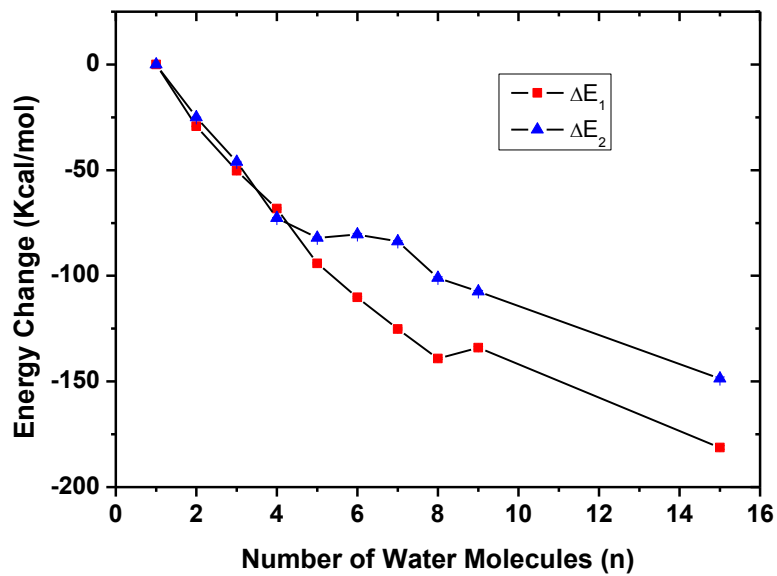

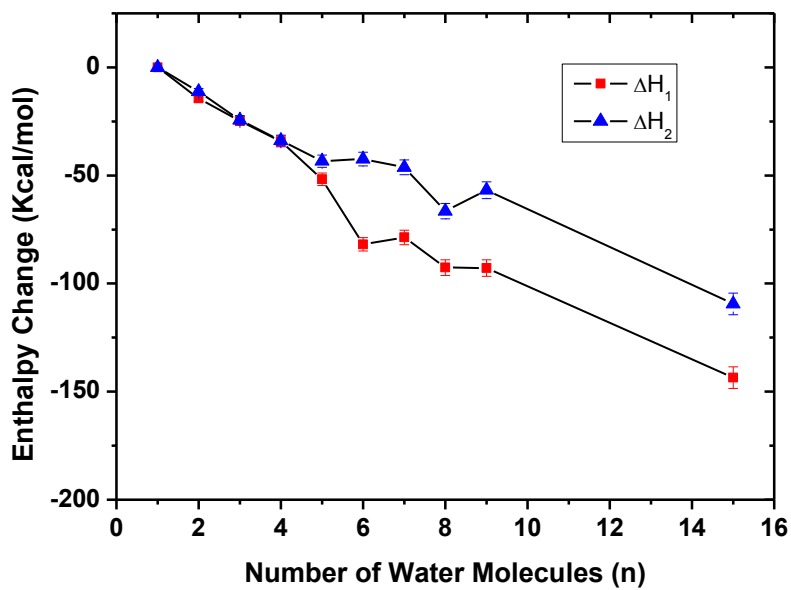

Fig. 4 (a)/(b) Variation of Energy/Enthalpy in system 1 and system 2 as a function of the water numbers. The standard deviation of energy is smaller than symbols, and the standard deviation of enthalpy is less than 5.0.

The energy/enthalpy changes in isolated S1 and S2, shown in Figure 4, are not hydration energies which are the amount of energy released when ions undergo hydrations, thus, attention is restricted to the relative energy difference between the two competing scenarios and $\Delta E_3$. According to Fig. 2, $\Delta E=\Delta E_1+\Delta E_2+\Delta E_3$ or $\Delta H=\Delta H_1+\Delta H_2+\Delta H_3$ where $\Delta E_3/\Delta H_3$ is the energy/enthalpy change in the reaction of $Na_2CO_3 + H_2O \Leftrightarrow NaHCO_3 + NaOH$ in vacuum. Enthalpy includes the total energy difference at ground state $\Delta E_{total}$ and finite temperature correction enthalpy difference $\Delta H_{total}$ between reactants and products, i.e. $\Delta H= \Delta E_{total} + \Delta H_{total}^{298.15K}$. The resulting energy and enthalpy are deduced as -10.377 kcal/mol and -10.191 kcal/mol respectively. The negative sign of the energy change indicates that this reaction can occur spontaneously at room temperature. Based on the above MM and QM energy calculations, the total energy change of reaction pathway Eq. (6), can be plotted as a function of the number of water molecules in Fig. 5. With less than 5 surrounding water molecules, the energy value is negative favoring the reaction pathway, and the negative energy value fluctuates with the conformation variation of hydration shells. However, when the number of water increases from 5 to 15 molecules, the hydration energy difference increases rapidly from negative to positive, then approaches a steady plateau of about 23 kcal/mol. The reason is that the effect of ions on water molecules becomes gradually smaller with more water molecules present, and approaches the bulk limit. This stands in contrast to the large impact on the average water molecule in the hydration shell when less water is available.

The variations and trend in Fig. 5 shows that with the reduction of the number of water molecules, it becomes more energetically favorable to form $HCO_3^-$ ion and $OH^-$ ion hydration in a relative dry condition, whereas forming $CO_3^{2-}$ ion hydration in relative wet condition. The $OH^-$ ions promote the absorption of $CO_2$ since $OH^-$ ions react with $CO_2$ in gas-phase without a free energy barrier.[53] the hydrolysis effect on $CO_3^{2-}$ ions increases with the reduction of the number of water molecules because ion hydration shells have a greater effect on the $CO_3^{2-}$ hydrolysis equilibrium constant than bulk water, which results in this counterintuitive phenomenon. This discovery sheds some light on the molecular mechanism of the observed phenomenon of dry absorption of $CO_2$. In what follows, several environmental factors affecting the moisture swing absorption of carbon dioxide are analyzed, which could enhance its system efficiency.

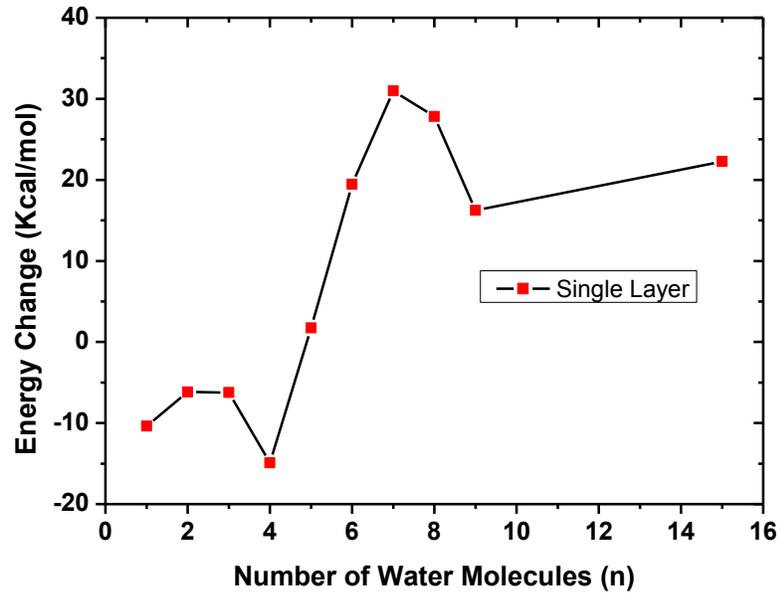

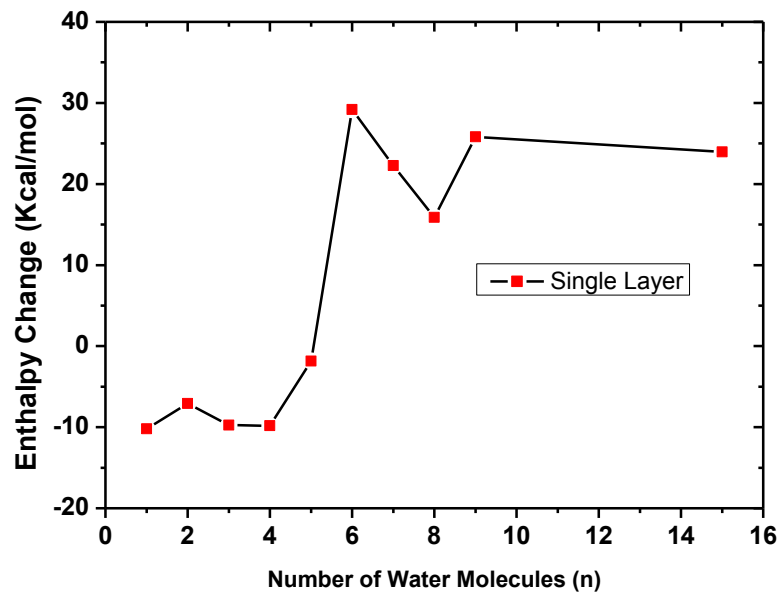

Fig. 5 (a)/(b) Eq. 6 Chemical reaction energy/enthalpy change with water numbers. $\Delta E_1/\Delta H_1$ and $\Delta E_2/\Delta H_2$ are the mean values shown in Fig. 4.

## 3.3 Parametric study of the moisture-driven $CO_2$ capture system

### 3.3.1 Effect of distance of confinement layers

The analyses above are based on the proof-of-concept model which consists of a mono-surface of graphene layer. The surface effect is now examined by exploring two competing systems: one confined between two layers and one "bulk system" without a surface. The former consists of ions and water molecules sandwiched between two parallel graphene layers with distance $D$ = 5Å, 7Å, and 9Å (three models) between them, and the latter consists of only unconstrained ions and water molecules in vacuum, shown in Fig. 6. Note that the $Na^+$ cations are still fixed in their respective spatial locations (with the same pattern 3.5 Å $\times$ 14 Å as that in the proof-of-concept model).

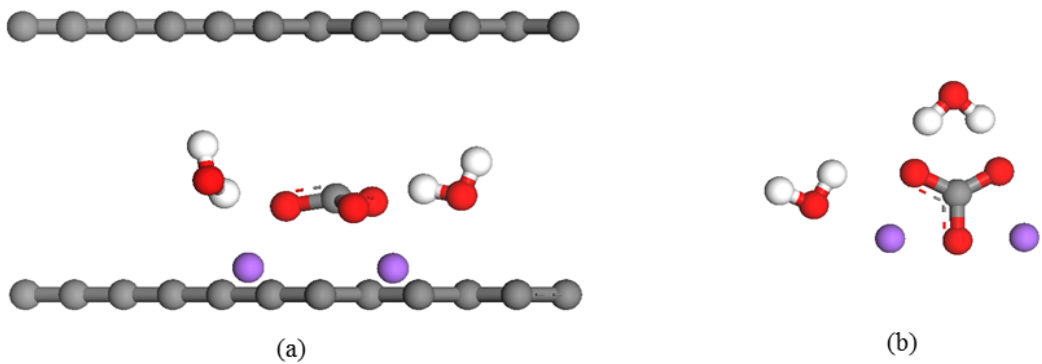

Fig. 6 (a) system confined between two graphene layers (b) bulk system

Following the same MD/QM simulation procedure, Fig. 7 plots the variation of the total energy/enthalpy change of the reaction in Eq. (6) as a function of water molecules in the system. The energy/enthalpy change of the chemical reaction without confined layers is positive, shown as black line; whereas that in the 5Å confined system is negative, shown as blue line. The results notify that the smaller distance between two confined layers is more favorable to forming $OH^-$ ions under the same humidity condition, which is more beneficial for absorbing $CO_2$ from the surrounding air. In essence, the confinement affects the geometry formation of ion hydrations and the hydrogen bonds: in the confined system, ion hydrations are physically enforced to become two-dimensional in form, whereas in the bulk system hydration shell formation is more complete. The smaller interlayer distance is more conducive to maintaining the two-dimensional shape of hydration layers. In this geometry configuration, the energy state of $OH^-$ and $HCO_3^-$ ion hydrations is more stable than the hydrated $CO_3^{2-}$ ion. When the distance between the interlayer is larger than 9Å, the impact of nanoscale confinement on the chemical reaction Eq.6 is same as the effect of one single surface layer. This indicates that the application of nanoporous materials may be attractive for absorbing $CO_2$, providing a feasible strategy of improving the efficiency of moisture-driven $CO_2$ capture system, in addition to the benefit of a large surface area (and hence more absorption sites).

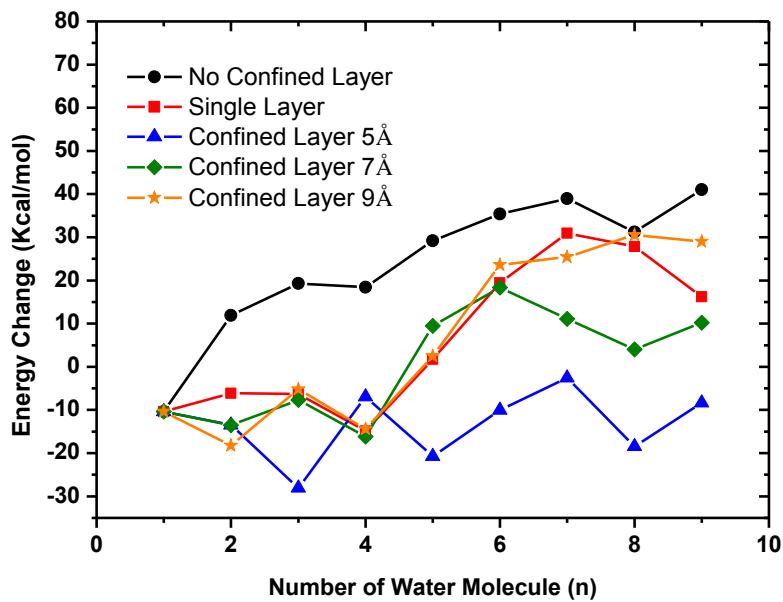

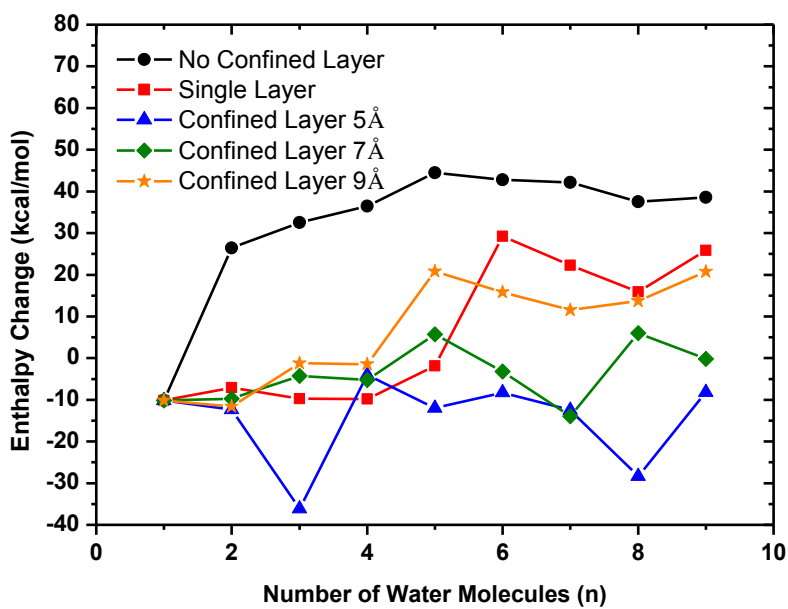

Fig. 7 (a)/(b) Eq. 6 Chemical reaction energy/enthalpy change with water numbers.

### 3.3.2 Effect of spacing of cations

The spacing between the cations on the solid surface is another key factor which has a pronounced effect on the absorption efficiency of moisture-driven $CO_2$ capture system. Fig. 8 shows the energy/enthalpy change of the reaction in Eq. 6, under three rectangular patterns of sodium cations with different spacings: 14 Å ×3.5 Å, 14 Å ×7 Å, and 14 Å ×14 Å, respectively. All systems are confined between two graphene layers with separation of 7 Å. The 14 Å ×3.5 Å rectangular pattern renders an obvious increase in the degree of chemical reaction in Eq. 6. In essence, the geometry configuration of the ion distribution has a decisive influence. When the distance of two $Na^+$ ions is relative close (3.5Å), a cross-shaped geometry configuration is formed by a $HCO_3^-$ and a $OH^-$ anion with the two $Na^+$ cations. Energy level of this geometry configuration is lower than the one of a $CO_3^{2-}$ ion locates in the middle of two $Na^+$ ions under the condition of small number of water molecules, so that the reaction product tends to be $HCO_3^-$ and $OH^-$ ions. However, when the distance of two $Na^+$ ions is relative far (7.0Å), $HCO_3^-$ and $OH^-$ anions are located in the vicinity of each $Na^+$ cation, energy level of this geometry configuration is higher than the one of a $CO_3^{2-}$ ion is in the middle of two $Na^+$ ions, which goes against the formation of $HCO_3^-$ and $OH^-$ ions.

In practice, the spacing or pattern of cations can be controlled by surface modification, such as attaching or self-assembling different groups of molecules on the surface, using nanoporous material with different pore size/structure with different load of carbonate ions, or adjusting the ion charge density on an ion exchange resin.

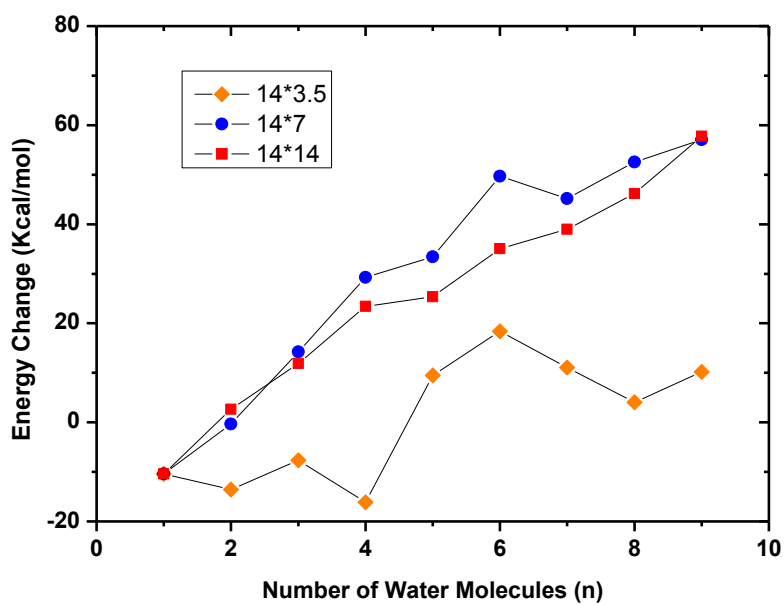

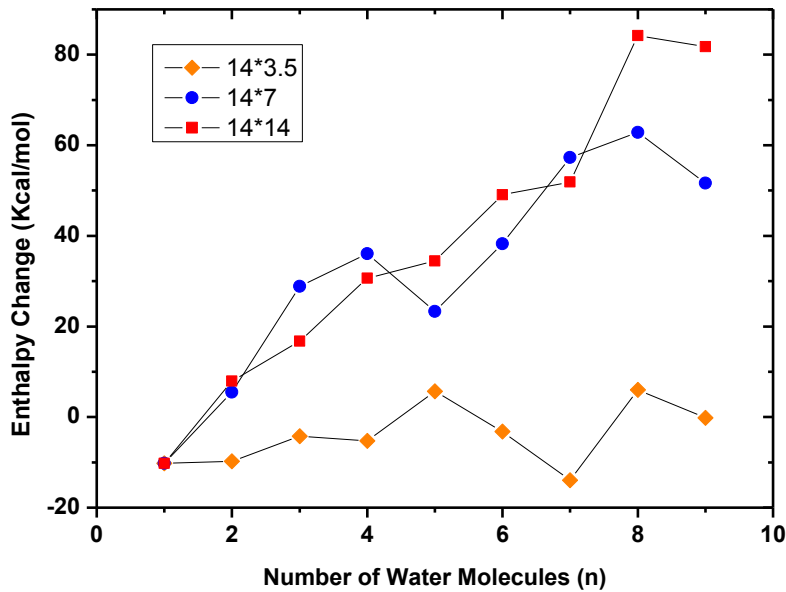

Fig 8. (a)/(b) Eq. 6 Chemical reaction energy/enthalpy change with water numbers.

### 3.3.3 Effect of surface treatment

Other factors can also affect the energy/enthalpy change of the reaction in Eq. (6). Besides surface confinement and cation distance explored in Section 3.3.1 and 3.3.2, surface modification is another one. Fig. 9 compares two $CO_2$ capture systems confined by two hydrophilic hydroxyl graphene layers and that sandwiched between two hydrophobic graphene layers. The distance of confined layers and the patterns of attached cations are identical in both systems. Fig. 10 shows that with the electrostatic attraction of the hydrogen bonds between water molecules and hydroxyls on hydrophilic surface, the hydrophilic layer is less conducive to generate $OH^-$ ions, and thus less welcoming the formation of $OH^-$ ions. Intrinsically, the solvation layers arise between two hydrophilic layers is not only as a result of the water molecules are physically confined between two surfaces and the existing anions and cations, but also as the hydrogen bonding between the water molecules and the hydroxyl surface,[54] wherefore the hydrophilic layer undermines the original 2-D geometry configuration of ion hydration formed by ions and hydrophobic confined layers.

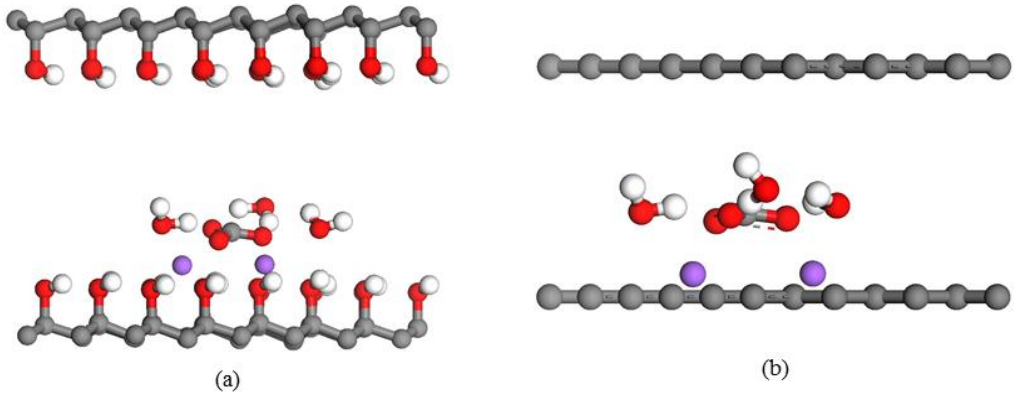

Fig. 9 Water-driven $CO_2$ capture system (a) Hydrophilic layer, partial charges of 0.412e and -0.57e are imposed on each hydrogen and oxygen atom of hydroxyl (b) Hydrophobic layer

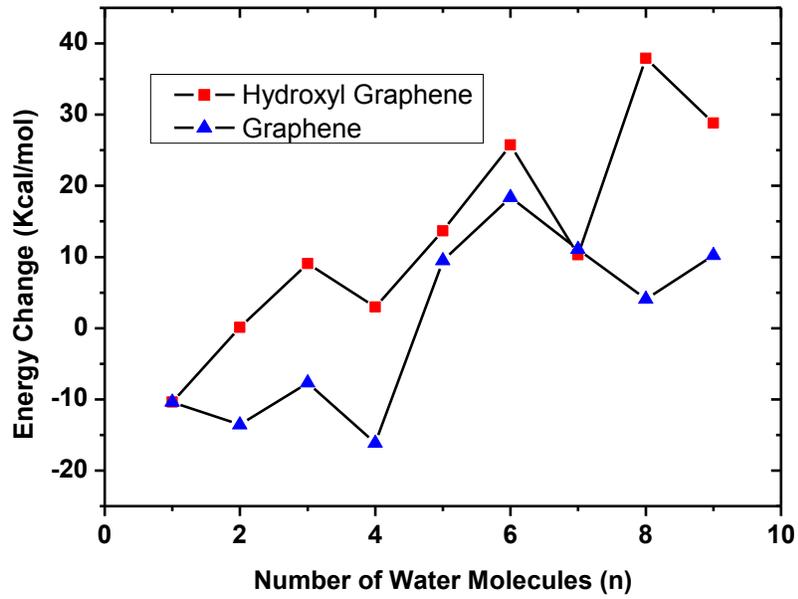

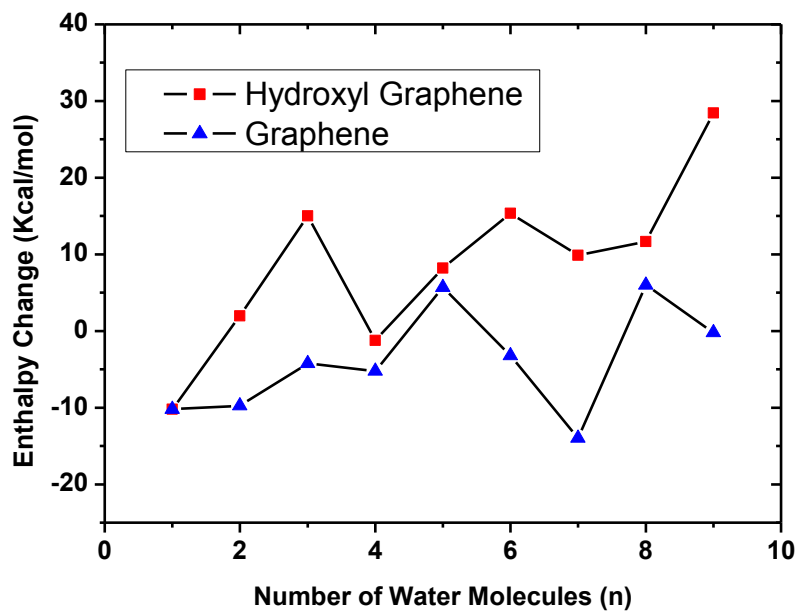

Fig 10. (a)/(b) Eq. 6 Chemical reaction energy/enthalpy change with water numbers.

### 3.3.4 Effect of temperature

Another factor governing system performance may be the ambient temperature. Fig. 11 shows that under the same humidity, the higher temperature is feasible to produce a larger amount of $OH^-$ ions, enhancing carbon dioxide absorption efficiency. The enthalpies of reactants and products both have increased because of the rise of temperature, the increased amounts are different leading to a slightly greater relative energy gap, which makes Eq.6 have more trends to react to the direction of product. The increased temperature may also help to increase the number of effective collisions between molecules to overcome energy barrier, and then produce more $OH^-$ ions.

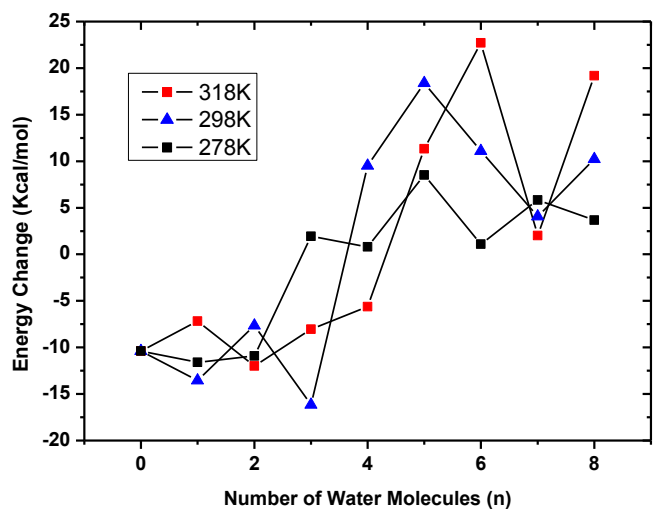

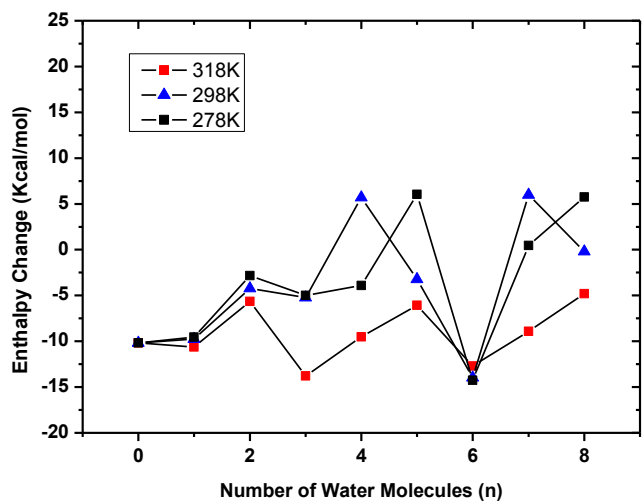

Fig 11. (a)/(b) Eq. 6 Chemical reaction energy/enthalpy change with water numbers at different temperatures

A more systematic parametric investigation may be carried out in future to optimize the material and system parameters, such as the confined pore size, cation's distance or pattern on solid surface, hydrophobic surface treatment, and the temperature, *etc*.

## 4. Experimental verification

Based on the molecular mechanism of moisture-driven $CO_2$ capture system, various nanoporous materials may be tested for absorption and desorption of carbon dioxide. Fig. 12, shows a number of examples of porous structures, whose performance may be modified using the aforementioned factors.

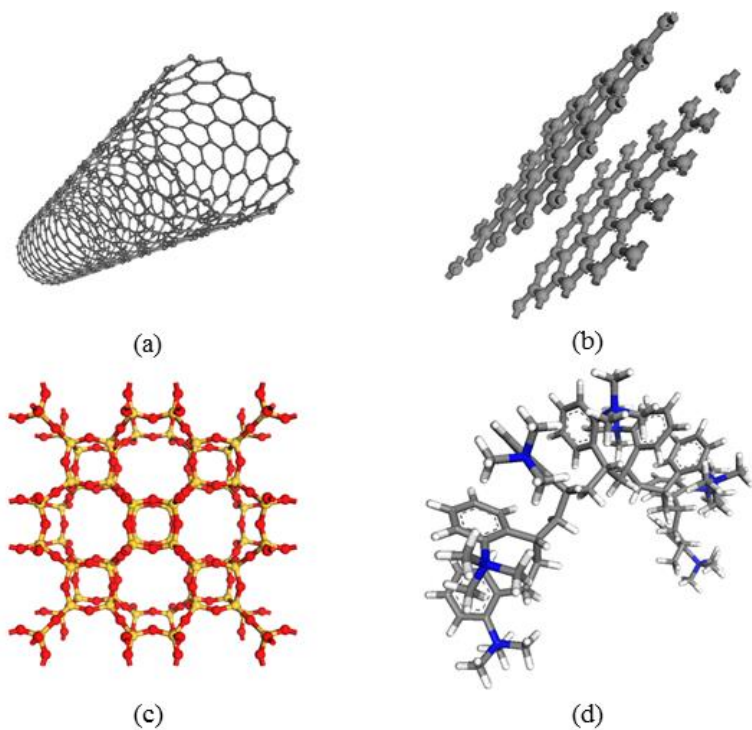

Fig. 12 (a) Carbon nanotube (b) Activated carbon (c) Zeolite (d) Ion exchange resin. Grey ball is carbon, red is oxygen, yellow is silicon, blue is nitrogen, and white is hydrogen.

### 4.1 $CO_2$ capture system driven by water quantities

To validate the feasibility of the water-driven $CO_2$ capture system, we performed a humidity controlling test on $CO_2$ capture based on an ion exchange resin (IER) material. The sorbent sample was prepared by soaking 0.30g IER into a 1M sodium carbonate solution for 4 hours, then dried out in ambient air. The prepared sample was put into a sealed chamber of an experimental device with temperature and humidity control. (see supplement material Fig. 1) The air humidity dew point in the experimental device was set first to 1.0°C and then turned down to 15°C to detect the variation of $CO_2$ concentration which was measured by an infrared gas analyzer (IRGA, LI-840)

The $CO_2$ concentration in the sample chamber changes with relative humidity. This is shown in Fig. 13. It clearly shows that IER with carbonate ion has a moisture effect on $CO_2$ absorption. The $CO_2$ absorption process takes place when the dew point is reduced. The lower dew point leads carbonate ion to hydrolyze larger amount of hydroxide ions to react with $CO_2$ in gas-phase without a free energy barrier[53]. However, if the sorbent is exposed to higher relative humidity level, the concentration of $CO_2$ equilibrates at a relative higher level because of smaller amount of hydroxide ions.

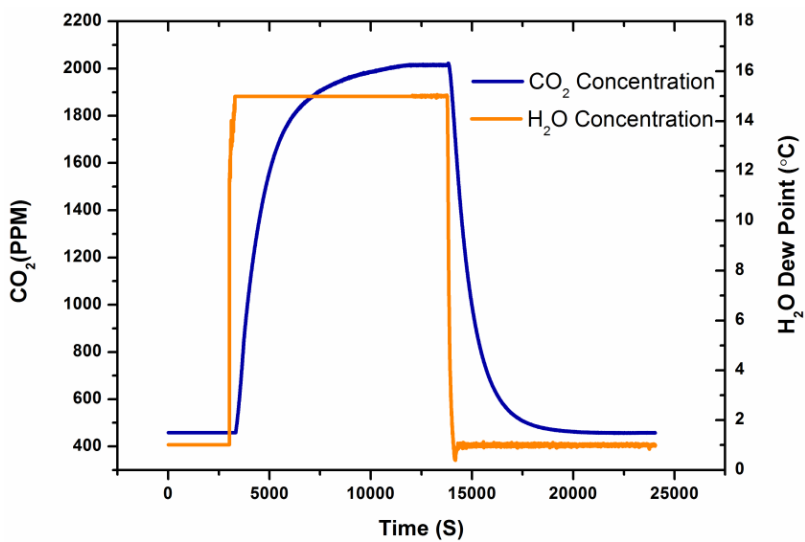

Fig. 13 Experimental result of $CO_2$ absorption curve at different humidity levels.

### 4.2 Effect of distance of confinement layers

In order to prove the nanoporous materials with confine-layer structure are attractive for absorbing $CO_2$ by producing larger amount of $OH^-$ ions than single-layer structure, four candidate samples: 1st candidate nanostructured graphite containing nanopores (~300nm) 0.1610g and 2nd candidate single-layer graphene 0.1494g were prepared by dripping 1M $Na_2CO_3$ solution 0.2cc on each sample, and then dried in vacuum chamber. The weights of ready-to-test samples are 0.1820g and 0.1689g, respectively, both carrying around 0.02g $Na_2CO_3$ powder. The 3$^{rd}$ candidate 0.02g pure $Na_2CO_3$ powder and 4$^{th}$ candidate 0.1610g nanostructured graphite were also prepared as reference. The same experimental device (see supplement material Fig. 1) was employed to determine the amount of $CO_2$ under different humidity conditions, shown as Figure 14. All four fresh samples were put into the experimental device at the same starting state: 655 ppm $CO_2$ concentration and 15℃ Dew Point water concentration. The equilibrium $CO_2$ concentrations of each sample at 15℃ Dew Point and 5℃ Dew Point were measured. The 1$^{st}$ sample nanomaterial with $Na_2CO_3$ shows a clear $CO_2$ concentration variation under different humidity condition, 655 ppm $CO_2$ concentration at 15℃ Dew Point and 450 ppm $CO_2$ concentration at 5℃ Dew Point. The 4$^{th}$ sample nanostructured graphite also has a minor variation of $CO_2$ concentration under different humidity conditions. This is a common phenomenon of physical adsorption. Sample adsorbs more water molecules when the water pressure above the sample increases while desorbs $CO_2$ leading $CO_2$ concentration increases. However, sample adsorbs less water molecules when the water pressure above the sample decreases while absorbs $CO_2$ leading to $CO_2$ concentration decreases. Note that single-layer graphene sample and pure $Na_2CO_3$ powder don't show the humidity swing which means these two absorbents cannot be regenerated by increasing water amount. This experiment verifies that the confined nanopores cause the moisture swing $CO_2$ sorbent to absorb $CO_2$ when surrounding is dry while release $CO_2$ when surrounding is wet.

Meanwhile, the $CO_2$ absorption capacity of the four candidates were measured under the same humidity condition at Dew Point 5°C, shown as Figure 15. The experiment results show that nanostructured graphite sample absorbed 2.80cc $CO_2$ which is more than 0.65cc, 0.50cc $CO_2$ and 0.45cc $CO_2$ absorbed by 2$^{nd}$ single-layer graphene sample, 3$^{rd}$ $Na_2CO_3$ powder and 4$^{th}$ nanostructured graphite sample, respectively. This experiment qualitatively verifies the theoretical results in Figure 7 and provides a feasible strategy of improving the efficiency of moisture-driven $CO_2$ sorbent. Capacities of sorbents with various pore sizes will be proceeded in the next step. The objective is to find the optimal pore size to enhance the capacity of moisture swing $CO_2$ sorbent.

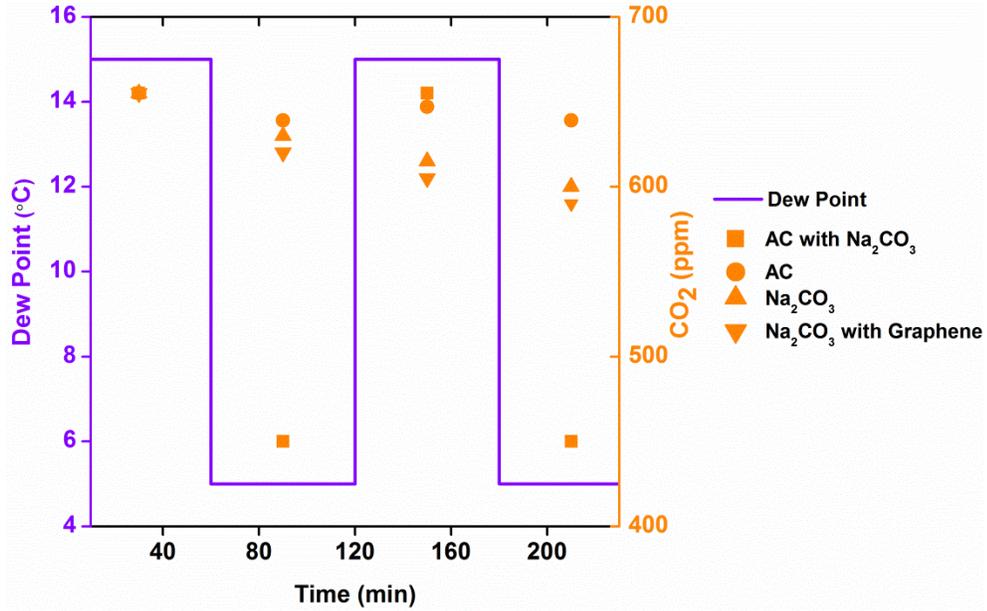

Fig. 14: $CO_2$ concentration change with different water numbers under the condition of different distance of confined layers.

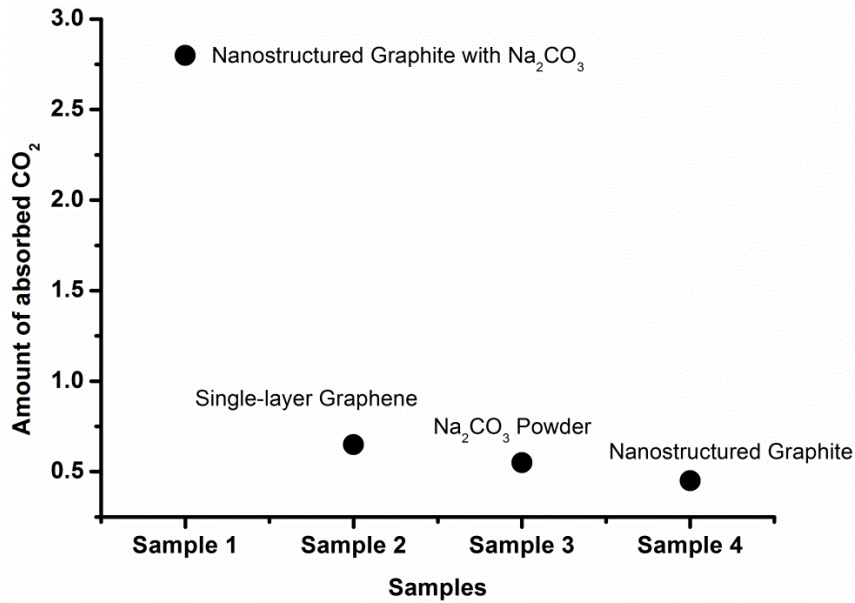

Fig. 15: $CO_2$ absorption capacity of four different samples. Sample 1 is Nanostructured Graphite with $Na_2CO_3$, Sample 2 is Single-layer Graphene, Sample 3 is $Na_2CO_3$ Powder, Sample 4 is Nanostructured Graphite.

The experiment about effect of distance between cations will be performed by using Ion Exchange Resin with different ion charge densities, and the temperature effect will be fulfilled in incubator next step.

**4.3 Effect of temperature**

In order to prove the higher temperature is feasible to produce a larger amount of $OH^-$ ions, enhancing $CO_2$ absorption efficiency, we performed a set of experiments to capture $CO_2$ by an ion exchange resin (IER) sample at the same humidity condition under different of temperatures. Another experimental device (see supplement material Fig. 2) was employed to maintain a constant humidity condition and different temperature levels. The concentrations of $CO_2$ in the device at different operating conditions are shown in Figure 16. The higher $CO_2$ concentration level means that water-driven $CO_2$ capture sample can absorb more $CO_2$ under the condition of higher temperature and the same humidity level, which leave the less amount of $CO_2$ in the air of experimental device.

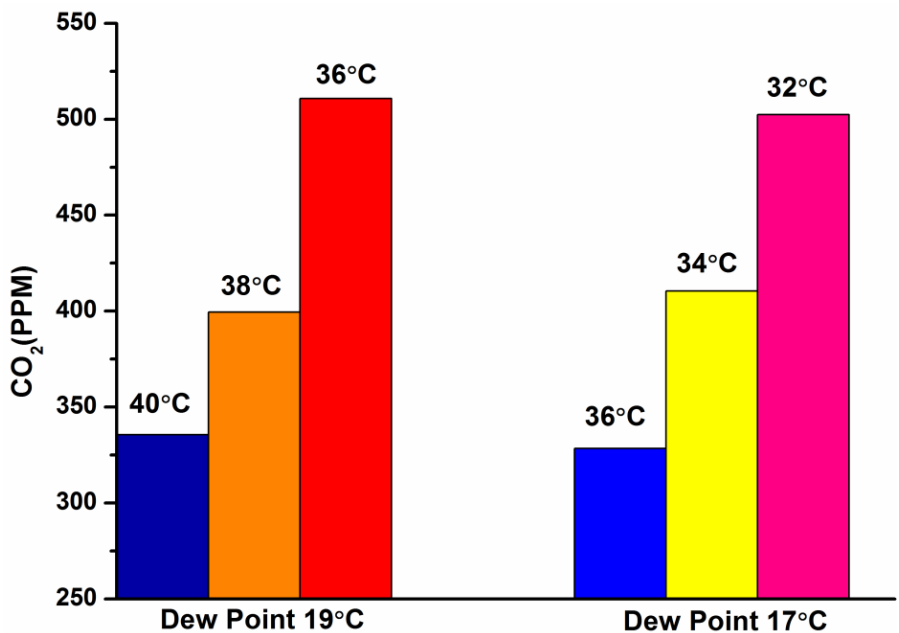

Fig. 16: $CO_2$ concentration in the experimental device with a water-driven $CO_2$ capture sample under two different humidity levels and different temperatures.

## 5. Concluding remarks

The change in energetically favorable states of different ion species with different water quantities underpins water-driven $CO_2$ capture system from ambient air. Using MD combined with QM simulations, the deduced hydration ion free energy shows that $CO_2$ capture system energetically prefers bicarbonate and hydroxide ion than carbonate ion when the environment is dry, and the resulting high content of hydroxide ion is more attractive for carbon dioxide absorption. Moreover, the effects of pore size, hydrophobic or hydrophilic confined layer, temperature, and distance of cations on the efficiency of $CO_2$ capture system are illustrated via the amount variation of hydroxide ions as the function of water quantity. A parallel $CO_2$ absorption experiment by ion exchange resin is carried out to verify the working principles and simulation findings.

The MD combined QM methodology developed in this paper provides a more efficient way to study similar problems which can be depicted by thermodynamic cycle as Fig. 2. The higher degree of the hydrolysis reaction between carbonate ion and water molecules at solid/water interface in a relative dry environment, may be applicable to other weak base and weak acid ions. This counterintuitive phenomenon also sheds some light on the fundamental interactions of ion hydrations in a confined space of solid materials. The underlying mechanism comprehension and parametric studies will help the design of more efficient and low-cost, energy-saving, humidity-driven $CO_2$ capture absorbent. More comprehensive optimization may be carried out in future.

## Acknowledgements:

This work is supported by a grant from the Office of Naval Research (ONR BAA 14-001). X.C. acknowledges the support from the National Natural Science Foundation of China (11172231 and 11372241), ARPA-E (DE-AR0000396) and AFOSR (FA9550-12-1-0159).

## Competing financial interests:

The authors declare no competing financial interests.